\documentclass[twocolumn]{aastex63} 
\usepackage{amsmath,  amssymb, url, graphicx}
\usepackage{revsymb}
\renewcommand{\vec}[1]{\pmb{#1}}

\newcommand{\beq}{\begin{equation}}
\newcommand{\eeq}{\end{equation}}

\newcommand{\T}{{\cal T}}

\newcommand{\M}{{\cal M}}

\def\Eq{Equation}
\def\Eqs{Equations}

\shorttitle{Hyperactive Magnetar Eruptions}

\begin{document}
\title{Hyperactive Magnetar Eruptions: Giant Flares, Baryon Ejections, and Fast Radio Bursts}

\correspondingauthor{Ashley Bransgrove}
\email{abransgrove@princeton.edu}

\author{Ashley Bransgrove} 
\affil{Princeton Center for Theoretical Science and Department of Astrophysical Sciences, Princeton University, Princeton, NJ 08544, USA}
\affil{Physics Department and Columbia Astrophysics Laboratory, Columbia University, 538 West 120th Street, New York, NY 10027}

\author{Andrei M. Beloborodov}
\affil{Physics Department and Columbia Astrophysics Laboratory, Columbia University, 538 West 120th Street, New York, NY 10027}
\affiliation{Max Planck Institute for Astrophysics, Karl-Schwarzschild-Str. 1, D-85741, Garching, Germany}

\author{Yuri Levin}
\affil{Physics Department and Columbia Astrophysics Laboratory, Columbia University, 538 West 120th Street, New York, NY 10027}
\affil{Center for Computational Astrophysics, Flatiron Institute, 162 5th Avenue, 6th floor, New York, NY 10010
}
\affil{Department of Physics and Astronomy, Monash University, Clayton, VIC 3800, Australia}




\begin{abstract}

Young neutron stars born with magnetic fields $B\gtrsim 10^{16}$\,G become hyperactive as the field inside the star evolves through ambipolar diffusion on a timescale $\sim 10^9$\,s. We simulate this process numerically and find that it can eject magnetic loops from the star. The internal magnetic field first diffuses to the crust surrounding the liquid core and then erupts from the surface, taking a significant amount of crustal material with it.
The eruption involves magnetic reconnection, generating a giant gamma-ray flare. A significant fraction of the eruption energy is carried by the neutron-rich crustal material, which must go through a phase of decompression and nuclear heating. The massive ejecta should produce additional emission components after the giant flare, including radioactively powered gamma-rays, optical emission, and much later a radio afterglow. The predicted eruptions may rarely happen in observed magnetars in our galaxy, which are relatively old and rarely produce giant flares. The model can, however, explain the extremely powerful flare from SGR 1806-20 in December 2004, its ejecta mass, and afterglow. More active, younger magnetars may produce frequent crustal eruptions and form unusual nebulae. Such hyperactive magnetars are candidates for the central engines of cosmological fast radio bursts (FRBs). We argue that each eruption launches an ultrarelativistic magnetosonic pulse leading the ejecta and steepening into a relativistic shock capable of emitting an FRB.
\end{abstract}


\section{Introduction}

The remarkable activity of magnetars is associated with the evolution of their ultra-strong magnetic fields \citep{duncan_formation_1992}. It generates persistent X-ray emission, bursts, and unusual changes in rotation rates \citep{kaspi_magnetars_2017}. Magnetars are also studied as likely sources of fast radio bursts (FRBs, \cite{Petroff_2019}). Fast ambipolar diffusion of fields $B\gtrsim 10^{16}$\,G inside young magnetars (age $t\lesssim 100$\,yr) can make them hyper-active, potentially explaining frequently repeating FRBs \citep{beloborodov_flaring_2017}.

How ambipolar diffusion generates magnetospheric bursts was not demonstrated with a first-principle simulation, so details of this process are uncertain. It is also unclear if it can reproduce the rich phenomenology of magnetar activity in X/gamma-rays and radio waves. One challenge is posed by giant gamma-ray flares, including the exceptionally bright flare from SGR~1806-20 on December 24, 2004 \citep{Hurley_2005, Palmer_2005}. Its luminosity peak exceeded $10^{47}$\,erg\,s$^{-1}$ for $\sim 0.5$\,s, followed by a 6-minute emission tail modulated with the magnetar rotation period $P=7.56$\,s. Furthermore, radio afterglow observations indicate that the flare ejected heavy crustal material of mass $M_{\rm ej}\sim  10^{25}-10^{27}\,$g with speed $v_{\rm ej}\sim c/2$ \citep{gaensler_expanding_2005,Granot_2006}. Similar crustal ejections were proposed to form unusual radio nebulae around repeating FRBs, in particular FRB~121102 \citep{beloborodov_flaring_2017}.

The massive crustal ejection from the neutron star of radius $R_\star\approx 12$\,km is a significant puzzle. The magnetospheric gamma-ray flare can ablate some material from the star surface \citep{thompson_soft_1995}, however detailed calculations give an ablated mass of only $\sim 10^{18}$\,g \citep{Demidov_2022}. Recently, \cite{Cehula_2024} proposed that $M_{\rm ej}\sim 10^{26}$\,g may be ejected by a crustal shock if pressure $p=10^{30}-10^{31}$\,erg\,cm$^{-3}$ is suddenly applied to the crust from the heated magnetosphere. The resulting hot ejecta is an interesting site of $r$-process. A larger $M_{\rm ej}\sim 10^{27}$\,g of radioactive material can also explain the delayed MeV gamma-rays detected at $\sim 10^3$\,s after the flare \citep{Mereghetti_2005,Patel_2025}. However, producing the ejecta by a crustal shock would require energy budget $\sim 3p R_\star^3$, which is $10^2-10^3$ times larger than the  observed flare.\footnote{Another problem of the crustal shock scenario is that the applied thermal pressure would need to be immediately removed to allow massive ejection through a rarefaction wave.}

This Letter describes a novel   mechanism for giant eruptions triggered by ambipolar diffusion inside magnetars. The proposed mechanism produces both massive ejecta and a gamma-ray flare. It is also capable of generating bright FRBs. Ambipolar diffusion, the driver of eruptions, is discussed in section~2. We consider magnetars with ultrastrong internal fields $B\gtrsim 10^{16}$\,G and discuss  their magnetic evolution in the liquid core and the crust surrounding the core. Section~3 describes a global simulation of ambipolar diffusion starting with a simple initial configuration and demonstrating the eruption. Observational implications are discussed in section~4.

\section{Hyperactive Magnetars}
Young neutron stars with internal magnetic fields $B>10^{16}$~G sustain 
high internal temperatures $T\sim 10^9$~K due to strong heating by ambipolar diffusion, which lasts $\sim 10^2$\,yr \citep{beloborodov_magnetar_2016}. At this temperature the star's outer layers with density $\rho< 10^{11}$~g~cm$^{-3}$ are in a fluid state, while the higher-density layers are crystallized \citep{chamel_physics_2008}. The crust surrounds the magnetized liquid core of the star. We will use a simplified model of the core composed of neutrons, protons, and electrons. We assume that its magnetic field exceeds the second critical field $H_{c2}\sim 10^{16}$~G, so that proton superconductivity is quenched. 

The evolution studied in this paper occurs much slower than thermal conduction, so the interior temperature profile is quasi-steady. Fast thermal conduction keeps the core and the lower crust at a uniform temperature $T(t)$, and significant temperature gradient is sustained only in the outermost layers with density $\rho<10^9$\,g\,cm$^{-3}$ \citep{Yakovlev_2004}.
The typical internal temperature $T\sim 10^9$\,K is likely above the critical temperature for neutron superfluidity in the core $T_\text{crit}\sim 10^8 - 10^9$~K \citep{potekhin_neutron_2015}, but below the critical temperature for neutron superfluidity in the deep crust $T_\text{crit}\sim 10^9 - 10^{10}$~K  \citep{chamel_physics_2008}. Therefore, we assume that the core neutrons are normal, while neutrons in the deep crust are superfluid. Hyperactivity of young magnetars is driven by the 
evolution of their internal magnetic fields. Below we describe the mechanisms of this evolution.

\subsection{Magnetic Field Evolution in the Core} 
\label{core_evolution}
The main process controlling the evolution rate $\partial_t\vec{B}$ in the core is ambipolar diffusion, which is the drift of the proton-electron plasma relative to neutrons \citep{goldreich_magnetic_1992, thompson_1996,beloborodov_magnetar_2016, Skiathas_2024}. 
The ambipolar drift is driven by the Ampere force $\vec{j}\times\vec{B}/c$, and opposed by proton-neutron friction. In addition, when the flow is compressive, pressure gradients are perturbed, and a deviation from chemical $\beta$-equilibrium appears in the core. This effect also opposes the drift. At the same time, charged-current weak interactions (in particular, murca reactions) tend to restore chemical equilibrium and erase the pressure gradients that inhibit the ambipolar flow. 

At the high temperatures of hyperactive magnetars, the weak interactions are fast and the departure from $\beta$-equilibrium remains small. The ambipolar drift velocity is then controlled by the balance between the Ampere force and proton-neutron friction --- the so-called ``friction-dominated" regime \citep{goldreich_magnetic_1992,beloborodov_magnetar_2016}. For ultra-strong magnetic fields $B>10^{16}$~G, heating by ambipolar diffusion is sufficient to offset neutrino cooling for $\sim 10^2$\,yr (see Figure~3 in \cite{beloborodov_magnetar_2016}), thus maintaining an extended period of high temperature and rapid magnetic field evolution (the so-called ``extended thermal balance phase").

Previous simulations of magnetic field evolution studied
two regimes: (i) High temperature limit where $p\leftrightarrow n$ conversions and $p$-$n$ collisions have very high rates \citep{Moraga_2024}. In this ``strong coupling" regime ambipolar diffusion is suppressed, and single-fluid magnetohydrodynamics may be used. (ii) Low temperature limit, where the increased mean free path for $p$-$n$ collisions enables significant plasma motion relative to neutrons, but the murca reactions needed for $p\leftrightarrow n$ conversions are extremely slow \citep{Castillo_2020,moraga_2025}. It was proposed that in these two limits the neutron fluid itself develops motions faster than the ambipolar drift \citep{ofengeim_fast_2018,Castillo_2020,moraga_2025,castillo_2025}.

However, neither of these extreme regimes holds during the magnetar evolution studied in this Letter.
The most significant 
evolution is enabled in the intermediate temperature regime of thermal balance, when heating by developed ambipolar diffusion offsets neutrino cooling by murca reactions \citep{beloborodov_magnetar_2016}. The self-regulated temperature $T\sim 10^9$\,K is then low enough for $p$-$n$ friction to allow significant ambipolar drift, and high enough for frequent murca $p\leftrightarrow n$ conversions, which assist the ambipolar drift.
Our results below show that the neutron fluid does move during this phase, but slowly compared to the ambipolar flow.

\subsection{Magnetic Field Evolution in the Crust}
\label{crust_evolution}

Two usual effects contributing to $\partial_t\vec{B}$ in the crust are ohmic diffusion and Hall drift. We argue below that both are of little significance for hyperactive magnetars. The ohmic timescale is $t_{\rm ohm}\sim 4\pi\sigma \ell^2/c^2$ where $\ell$ is the scale of field variation and $\sigma$ is an effective conductivity. 
For relevant scales $\ell\gg 10^4$\,cm, the ohmic timescale exceeds the lifetime of magnetar hyperactivity. The timescale for Hall drift is $t_{\rm H}=e n\ell/j_\perp$, where $n$ is the electron density and $\vec{j}_\perp$ is the electric current component perpendicular to $\vec{B}$. This gives $t_{\rm H}\sim (4\pi e n\ell^2/cB)(j/j_\perp)$, which exceeds the lifetime of hyperactive magnetars for large-scale fields in the deep crust\footnote{Gradients of $\vec{B}$ can become significant at the onset of the eruption when the magnetic loop pinches off in the outer crust. The current density is large in the reconnection layer, and the Hall timescale then becomes much shorter. The effect of Hall drift on the crustal eruptions should be investigated in future simulations.
}.

In a magnetar with $B>10^{16}$\,G, the magnetic tension $B^2/4\pi$
dominates the elastic shear modulus $\mu$ in the entire crust:\footnote{A maximum shear modulus (at the bottom of the crust) is $\mu\approx 10^{30}$~erg~cm$^{-3}$ while  $B^2 /4\pi \approx 8\times 10^{30}\,B_{16}^2$\,erg\,cm$^{-3}$.
}
$B^2/4\pi\mu\gg 1$. 
The crust then effectively behaves like a magnetohydrodynamic fluid, even though the crystal is below its melting temperature, $T<T_\text{melt}$. In our axisymmetric model there are no gradient forces to balance the azimuthal component of the Ampere force, so $(\vec{j}\times\vec{B})_\phi \approx  0$ and the poloidal components of $\vec{j}$ and $\vec{B}$ are nearly parallel \citep{bransgrove_magnetic_2018}. This implies a suppression of Hall drift of the poloidal magnetic flux surfaces.

Instead of the usually considered Hall and ohmic effects, we find a new dominant mechanism for magnetic field evolution in the deep crust: the Ampere force induces a flow of the crustal plasma (lattice+electrons) through the ocean of superfluid neutrons, while nuclear reactions adjust the local composition. This mechanism is essentially a type of ambipolar diffusion. It operates in the deep crusts of neutron stars with $B\gtrsim 10^{16}$\,G. 

Ambipolar diffusion in the deep crust ($\rho>\rho_\text{drip}\approx 4\times 10^{11}$\,g\,cm$^{-3}$) proceeds similarly to that in the core. Magnetically induced compressions/decompressions of the crystal raise/lower the electron Fermi energy, and drive electron captures/$\beta$-decays of the lattice nuclei (Appendix~\ref{beta_reactions}, see also \cite{Chamel_2021, Chamel_2022}). When $\rho>\rho_\text{drip}$, 
the extra neutron resulting from electron capture is not kept by the nucleus but quickly emitted
\citep{Haensel_1990}. 
We assume that the emitted neutrons quickly down-scatter in the lattice and join the superfluid that freely flows through the crystal. 
Thus, compression leads to the loss of nucleons from the compressed region of the lattice, leaving behind superfluid neutrons. Despite the high electric conductivity, the magnetic flux is not frozen in the crust, allowing ambipolar flow. At the same time, the depletion of electrons reduces the pressure in the compressed region, enabling the ambipolar drift   (Appendix~\ref{beta_reactions}). Note also that shear plastic flows of the crust (relaxing the solenoidal Ampere forces) do not involve compression, so do not change the electron Fermi energy and do not induce $\beta$-reactions.\footnote{During radial motions of the lattice neutrons are emitted or absorbed by nuclei on very short timescales in order to preserve their chemical equilibrium.} 

At densities $\rho<\rho_\text{drip}$, electron captures are not followed by neutron emission. They reduce pressure in the lattice and facilitate slow compression, but do not violate the flux freezing condition and do not enable ambipolar diffusion.

\section{Numerical simulation}

\label{numerical}

\subsection{Method}
\label{numerical_setup}

In our simulation, the problem is reduced to its simplest form, with matter in the entire star modeled as a mixture of $npe$, with magnetic field $\vec{B}$ frozen in the plasma component ($e$ and $p$ move with equal velocities). The star has four regions defined by density $\rho$ and  $\beta=8\pi P/B^2$ ($P$ is the local hydrostatic pressure):
\begin{itemize}
\item
``Core" ($\rho>\rho_{\rm core}\equiv 10^{14}$~g~cm$^{-3}$): $n$ and $p$ components are assumed to be normal, not superfluid, and experiencing significant friction in response to their relative motion. 
\item 
``Lower crust" ($\rho_{\rm drip}<\rho<\rho_{\rm core}$): all neutrons are superfluid, with no friction against the plasma. 
\item 
``Upper crust" ($\rho<\rho_{\rm drip}$ and $\beta>0.1$): $npe$ are coupled and move as a single fluid.
\item 
``Magnetosphere" ($\beta<0.1$): $npe$ are coupled, and the magnetic field is kept near force-free balance. 
\end{itemize}
The equation of state for each component $e$, $p$, and $n$ is that of an ideal degenerate Fermi gas. Each component has its chemical potential and pressure determined by its number density.
We choose the simplest equation of state because we are not interested in the accurate density structure of the neutron star. Instead, we focus on qualitative features of ambipolar diffusion, which are easier to study with the simplified model. Conversions $e+p\leftrightarrow n$ (beta-reactions) are driven by deviations $\Delta\mu$ from chemical equilibrium between $e$, $p$, and $n$, which appear during the evolution (Appendix~\ref{setup}).

Our treatment of the lower crust is particularly crude, as we do not track its true nuclear composition. This turns out possible when beta-reactions are sufficiently fast, making the small value of $\Delta\mu$ (which depends on nuclear composition) unimportant for the magnetic field evolution, so any toy model for the small $\Delta\mu$ may be used. This aspect of our simulation is further described in Appendices~\ref{setup} and \ref{beta_reactions}, where we argue that the lower crust stays near $\beta$-equilibrium with superfluid neutrons, which easily flow through the plasma (with no friction) to maintain their preferred density distribution in the hydrostatic balance.

The rate of global magnetic evolution is then controlled by processes in the core. Here, we use the same $p$-$n$ friction, rate of $e+p\leftrightarrow n$ conversions, and thermal evolution equation as in \cite{beloborodov_magnetar_2016}. The thermal equation includes neutrino cooling and ambipolar heating (Appendix~\ref{ambipolar_appendix}). The system evolves slowly, except the final brief phase of dynamical eruption. The timescale of ambipolar diffusion in the core, $t_{\rm amb}\sim 100$~yr, is much longer than the Alfv\'en crossing time of the star, $t_{\rm A}\sim 10^{-2}$~s. So, the evolution proceeds through the sequence of magnetic two-fluid equilibria, which determines the instantaneous velocities of the plasma and the neutron fluid throughout the star. Our simulation method tracks this equilibrium sequence using the relaxation technique described in Appendix~\ref{setup}, which effectively filters out dynamical waves. Our numerical method also allows us to approximately track the final eruption phase, but in slow motion, with an artificially reduced acceleration, as we artificially reduce the developing uncompensated magnetic forces.

The set of equations governing the evolution are given in Appendix~\ref{ambipolar_appendix}, and the numerical scheme is described in Appendix~\ref{appendix_numerical}.
The simulation is axisymmetric. Its domain is defined in spherical coordinates $r$ and $\theta$, and covers $r \in [0,100]$~km and $\theta \in [0,\pi]$ with $N_r \times N_\theta = 350 \times 240 $ cells. The grid is uniform in $\theta$ and non-uniform in $r$, with resolution highly concentrated near $r_\star\sim 10$~km to resolve the scale-height in the outer crust. Outside the star, the radial grid smoothly transitions to log-radial spacing, so that the outer magnetosphere is captured with fewer cells. 

The initial condition is described in Appendix~\ref{appendix_numerical}. It consists of a poloidal magnetic field with dipole moment $\mu \approx 2.8\times 10^{34}$~G~cm$^3$, and 
a `twisted torus' localized inside the star. 
The initial temperature of the star is $T=3\times 10^9$~K. The fluid is initialized in an unmagnetized hydrostatic equilibrium. Since this is not an ideal MHD equilibrium, we allow the star to relax into an ideal MHD equilibrium configuration at fixed temperature, before the ambipolar diffusion simulation begins. After relaxing the star into MHD equilibrium, the poloidal field strength at the pole is $B\approx 4\times 10^{16}$~G. The internal toroidal field has maximum strength $B_{\phi,\text{max}} \approx 5\times 10^{16}$~G, and the rms internal field strength is $\bar{B} \approx  7\times 10^{16}$\,G. The total magnetic energy inside the star is $E_B \approx 10^{51}$~erg.  

\begin{figure}[t!]
\centering
\includegraphics[width=0.47\textwidth]{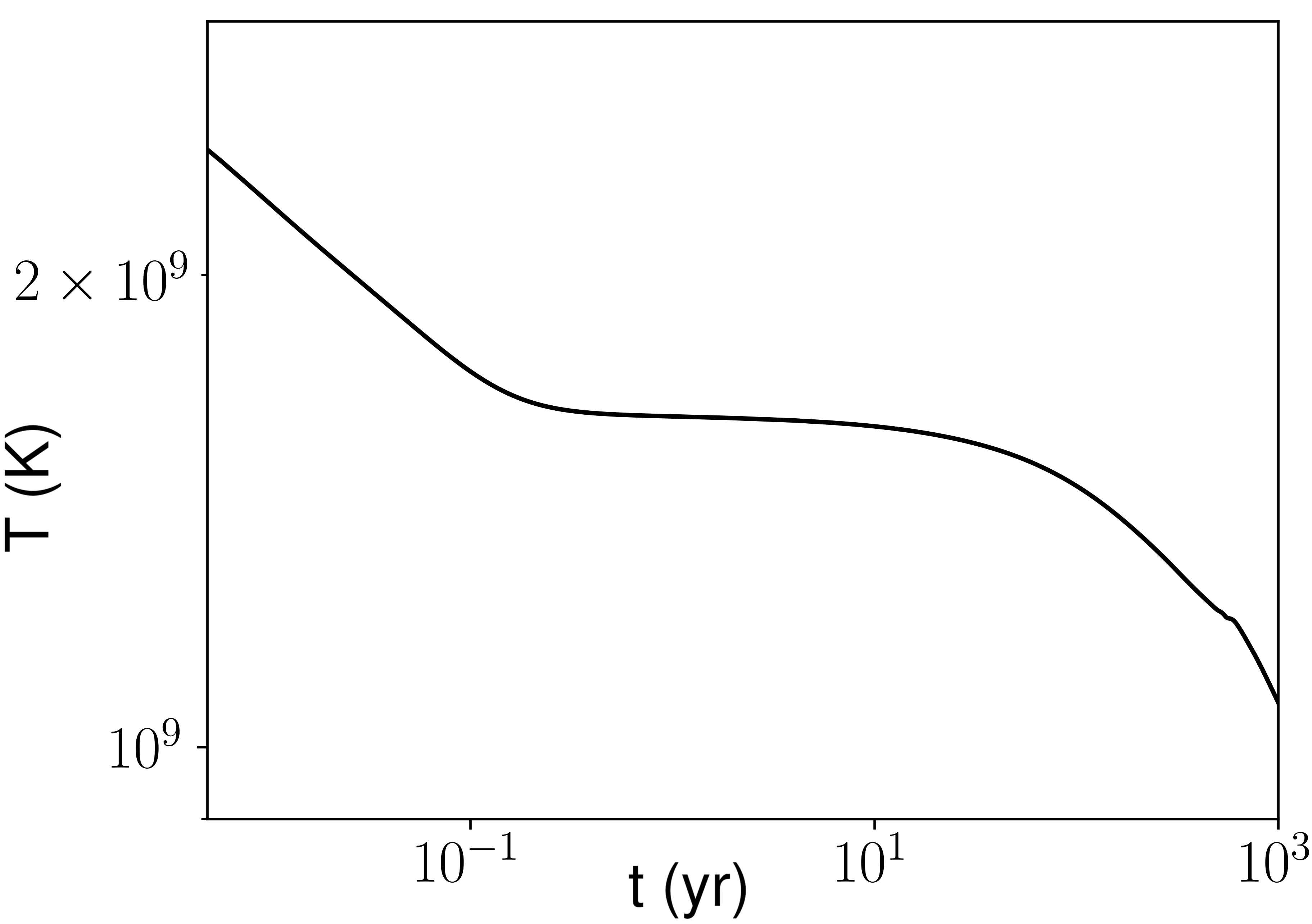} 
    \caption{Thermal evolution of the hyperactive magnetar. At early times the temperature drops rapidly due to neutrino cooling (ambipolar heating is inefficient at 
    high-$T$ 
    because collisional coupling inhibits plasma motion relative to neutrons).
    At $0.1$~yr~$\lesssim t \lesssim 400$~yr ambipolar heating balances neutrino cooling so that $T\approx$~const. At late times,
    the magnetic energy available for dissipation decreases, so ambipolar heating declines, and neutrino cooling again dominates.
    }
    \label{temperature}
\end{figure}

\subsection{Results}
\label{simulation}

Before examining the magnetic field evolution and fluid motions in the simulation, it is useful to look at an important global characteristic: the interior temperature of the neutron star $T(t)$ (Fig.~\ref{temperature}). At the beginning of the simulation,
$T> 2\times 10^9$~K ensures that protons and neutrons in the core are strongly coupled by collisions, and significant relative motions (ambipolar diffusion) do not occur. As a result, frictional heating is inefficient and the temperature drops rapidly due to neutrino cooling. When the star cools to $T\lesssim 2\times 10^9$~K, 
the neutrino cooling is reduced and the mean free path for $p$-$n$ collisions is increased, enabling significant ambipolar diffusion.

\begin{figure*}[t!]
\centering
\includegraphics[width=1.0\textwidth]{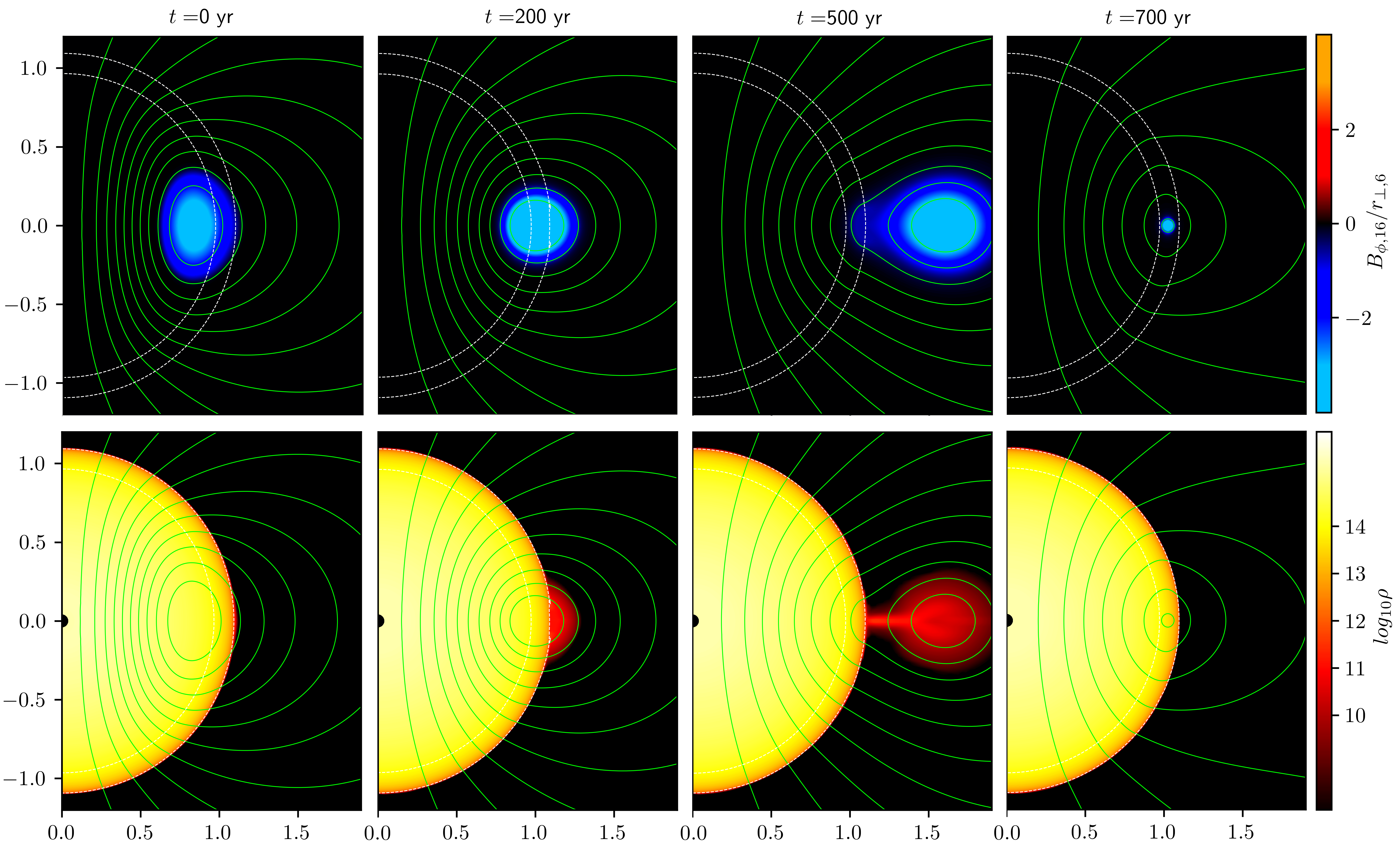} 
    \caption{Magnetic field evolution of the hyperactive magnetar. Four snapshots are shown at times $t=0$, 200\,yr, 500\,yr, and 700\,yr. Green curves represent poloidal magnetic field lines.
    The top row shows $B_\phi/(r\sin\theta)$. The bottom row shows log  of baryon density $\rho$ in units of g\,cm$^{-3}$. In each panel, the horizontal $x$-axis shows the distance from the vertical $z$-axis (the axis of symmetry); both $x$ and $z$ are in units of 10\,km.
    The inner and outer dashed white curves are 
    contours of $\rho=\rho_\text{core}$ and $\rho=\rho_\text{drip}$. 
    An animation of this figure is available in the online version of the article.}
    \label{evolution}
\end{figure*}

\begin{figure*}[t!]
\centering
\includegraphics[width=1.0\textwidth]{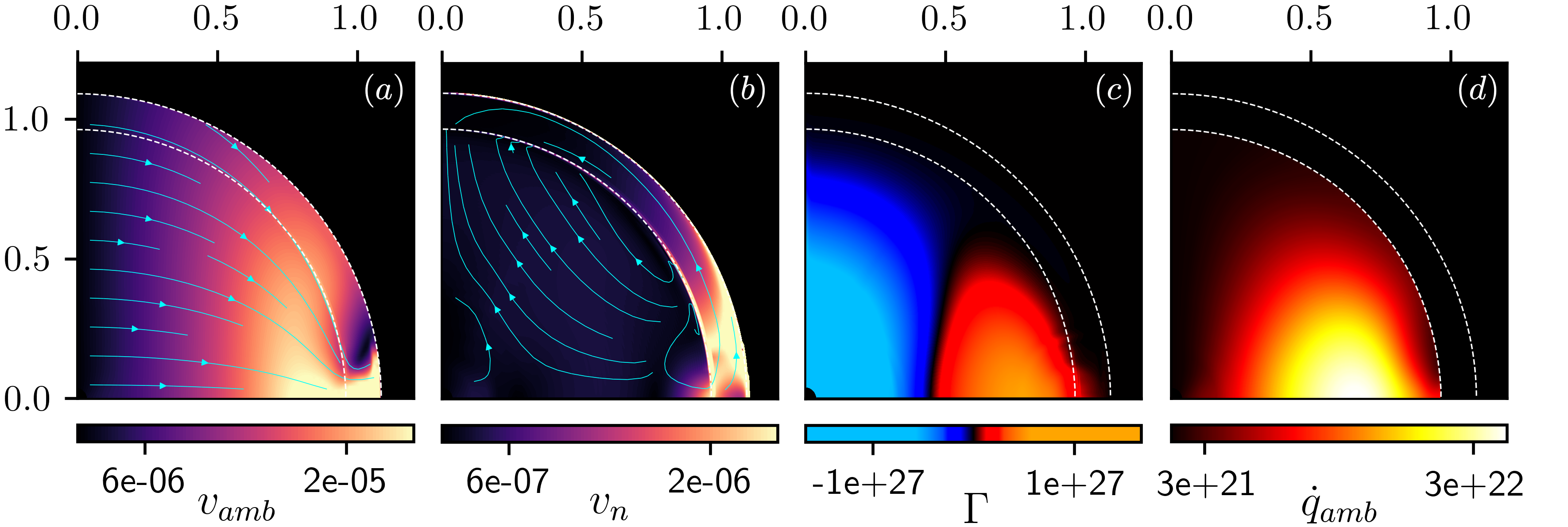} 
    \caption{
    Snapshot of ambipolar diffusion at $t=200$~yr. 
    (a) Relative velocity between plasma and neutrons (ambipolar diffusion velocity) $\vec{v}_{amb}$ in units cm~s$^{-1}$. (b) Neutron velocity $\vec{v}_n$ in units cm~s$^{-1}$. (c)
    Rate of $\beta$-reactions $\Gamma$ in units s$^{-1}$~cm$^{-3}$ (red 
    corresponds to
    neutron production, and blue 
    corresponds to
    electron-proton production). 
    (d) Ambipolar
    heating rate $\dot{q}_{amb}$ in units erg~s$^{-1}$~cm$^{-3}$. 
    Axes 
    are the same as Fig.~1.
    The dashed white curves are 
    the contours of $\rho=\rho_\text{core}$ and $\rho=\rho_\text{drip}$.
    An animation of this figure is available in the online version of the article.  } 
    \label{still}
\end{figure*}

After $\sim 1$~month, frictional heating becomes sufficient to offset neutrino cooling, marking the beginning of the extended `thermal balance' phase when the temperature $T = T_\text{bal}\approx 1.5\times 10^9$~K is nearly constant in time (Fig.~\ref{temperature}). Both $T_{\rm bal}$ and the duration of this phase are in agreement with \cite{beloborodov_magnetar_2016}.
$T_\text{bal}$ is low enough to permit significant relative motion between protons and neutrons, but high enough for modified urca reactions to quickly erase deviations from chemical $\beta$-equilibrium and maintain $|\Delta \mu| \ll kT$ throughout the thermal balance phase. 
The magnetic field strongly evolves during the thermal balance phase
on a timescale $t_{\rm amb}\sim 10^9$~s. 

The magnetic field evolution in the simulation is shown in Fig.~\ref{evolution}. It starts from the ideal MHD equilibrium with the initially strongly coupled plasma and neutrons (Section~\ref{numerical_setup}). As the star cools, the force balance changes because the plasma is now allowed to slowly move through the neutron fluid, so the system enters two-fluid evolution. The developing motion of plasma (with the frozen magnetic field lines) generates frictional heat. A snapshot of the two-fluid motions at a later phase is shown in Fig.~\ref{still}. It demonstrates the basic pattern of the plasma drift driven by magnetic stresses and the accompanying slow neutron flow induced in the two-fluid system.

As one can see in Fig.~\ref{evolution} and \ref{still}, the poloidal magnetic loops in the core contract under their self-tension force, converging toward the equator. The compression drives $\beta$-reactions that deplete electrons and produce neutrons in the outer core near the equator (the conversion rate is shown in Fig.~\ref{still}, panel c). The created neutrons diffuse out of the compressed region in order to maintain force balance. Note that fast $\beta$-reactions enable compressive flows that do not change the density profile of the star: $\nabla\cdot(n_p \vec{v}_p + n_n \vec{v}_n)\approx0$. This implies $v_n\sim v_p (n_p/n_n)\ll v_{\rm amb}$, 
as demonstrated by the simulation
(Fig.~\ref{still}).

On a timescale $\gtrsim 100$\,yr, the magnetic loop containing toroidal flux (blue region in Fig.~\ref{evolution}) is pushed toward the star's surface. This outward drift of the loop is driven by the magnetic pressure gradient. It proceeds unimpeded in the lower crust because of the fast $\beta$-reactions. Similarly to the dynamics in the core,  any compression of the plasma in the lower crust raises the local electron Fermi level, driving electron captures $e+p\rightarrow n$. The depletion of electrons prevents  any significant build-up of the plasma pressure in response to the magnetic field evolution, which enables the observed evolution. The excess of neutrons released in this process is easily redistributed toward the poles to maintain the hydrostatic profile of the star (Fig.~\ref{still} panel b), as neutrons are superfluid \citep{Gusakov_2020}.

As the magnetic loop diffuses outward, it pushes and lifts the outer crustal material with $\rho\lesssim \rho_\text{drip}$. This material does not support ambipolar diffusion and remains trapped in the magnetic field; however, its weight and pressure are incapable of arresting the magnetic field evolution. The magnetic loop loaded with crustal matter slowly emerges from the surface (Fig.~\ref{evolution}) and eventually pinches off. Just before the eruption, it contains magnetic energy $E_\text{flare}\approx 5\times 10^{49}$~erg and mass $M_\text{ej}\approx 4\times 10^{28}$~g. Its gravitational binding energy is $E_b \approx 0.1 M_\text{ej}c^2 \sim 4\times 10^{48}$~erg, and its average mass density is $\bar{\rho} \equiv M_\text{ej}/V\approx 0.1 \rho_\text{drip}$.
The eruption of this gigantic blob
occurs suddenly when the bottom part of the loop reaches the layer where $\beta\sim 1$ (nearly the entire loop at this moment is in the region of $\beta<1$). In our simulation, the magnetospheric eruption is artificially slowed and driven by a small uncompensated Ampere force; in reality it occurs on a dynamical Alfv\'en timescale (this can be captured in future full MHD simulations).
After the eruption, the star relaxes into a new equilibrium. At late times, ambipolar heating becomes inefficient because a significant fraction of the magnetic energy has been dissipated (Fig.~\ref{temperature}). 

\begin{figure}[t!]
\centering
\includegraphics[width=0.49\textwidth]{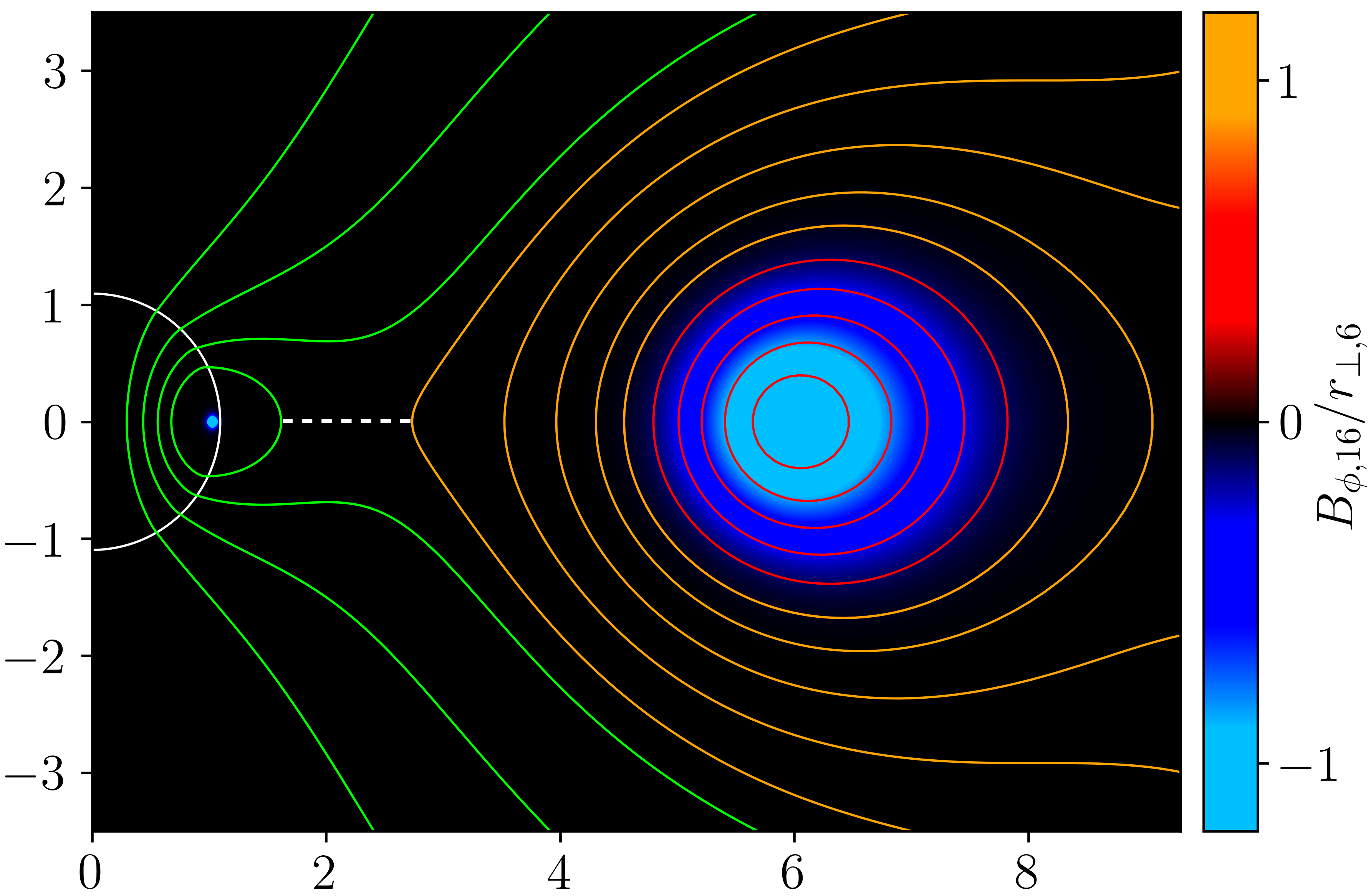} 
    \caption{Structure of the magnetospheric flare. Color 
    shows
    $B_\phi / r_\perp$, and the white curve is the neutron star surface. 
    Green curves 
    show
    field lines that are connected to the star. Red curves represent baryon-loaded field lines that were ejected from the star, and orange curves represent baryon-free field lines that reconnected around the ejecta. The dashed white line shows the location of the current sheet.}
    \label{magnetosphere}
\end{figure}

Although our simulation follows the dynamical eruption phase in ``slow motion,'' it does demonstrate basic features of the ejection process. In particular, it shows the change of magnetic topology via reconnection of field lines closing behind the ejected  plasmoid. As the heavy plasmoid escapes the magnetosphere, baryon-free field lines that were previously external to the matter-loaded loop must close around it and reconnect. Therefore, the ejected heavy baryonic blob becomes dressed in baryon-free field lines (Fig.~\ref{magnetosphere}, orange curves).

\section{
Observational implications
}

\subsection{Giant Gamma-ray Flare}
As the magnetic loop emerges from the star, it stores free energy in the magnetosphere comparable to the energy of the loop itself. Therefore, the magnetospheric free energy available for the flare is $E_\text{flare}\sim VB^2/8\pi$, where $V$ is the initial volume of the ejected loop. When the loop detaches, the current sheet that forms behind the ejection (Fig.~\ref{magnetosphere}, dashed line) is unpolluted by baryons.\footnote{In fully three-dimensional (non-axisymmetric) simulations, some baryons could leak from the ejected blob, which can be studied in future work.}
The current sheet dissipates energy $E_\gamma \sim 0.1 E_\text{flare}$ through magnetic reconnection on a timescale $\sim 10r/c$ \citep{Parfrey_2013}, with immediate heat reprocessing into a thermalized mixture of photons and electron-positron pairs. As a result of the reconnection process, the ejected heavy blob becomes dressed in baryon-free, hot, magnetic loops (Fig.~\ref{magnetosphere}, orange lines). The temperature of this dress
may be roughly estimated as $aT^4\sim E_{\rm flare}/r^3$ (where $a$ is the radiation constant), which gives $T$ in the MeV range.

A full magnetohydrodynamic simulation will be needed to investigate the acceleration of the dressed blob.
Note that the dress is puffy and magnetically dominated; its effective sound speed is close to $c$, so the dress would expand with speed $c$ if it did not embed the heavy baryonic blob.
Its dynamics is governed by magnetic stresses and radiation pressure, and its magnetic field lines may eventually stretch ahead of the slower blob and reconnect away to form a leading baryon-free outflow expanding freely with the speed of light. 

As the dress temperature drops due to adiabatic cooling, the ratio of $e^\pm$ and photon densities becomes exponentially suppressed, $n_\pm/n_\gamma\sim \exp(-m_e c^2/kT)$. Suppression by a few tens of $e$-foldings is sufficient to reach transparency and release the photons, producing a powerful burst. A similar release is known in models of cosmological gamma-ray bursts  \citep{Paczynski_1986}; it occurs when adiabatic cooling reduces the plasma temperature in its rest frame to $\sim 20\,{\rm keV}$. The resulting emission spectrum depends on the dress dynamics. The released radiation will be in the gamma-ray band if part of the hot dress does detach from the blob and freely accelerates outward to a Lorentz factor $\Gamma\gg 1$. Then, the Doppler shift by $\Gamma$ will give the observed radiation temperature $\sim 20\,\Gamma$~keV, not much below the initial MeV temperature of the dress at its formation in the reconnection layer.

Our simulation suggests a simple approximate relation between the ejected mass $M_{\rm ej}$ and the flare energy $E_{\rm flare}$. The characteristic density in the ejected blob at the onset of the eruption, $\overline{\rho}\sim 0.1 \rho_{\rm drip}$, corresponds to the blob volume $V\sim M_{\rm ej}/\overline{\rho}$ and $E_{\rm flare}\sim (M_{\rm ej}/\overline{\rho})(B^2/8\pi)$, which gives
\begin{equation}
  E_{\rm flare}\sim 10^{47}\,B_{16}^2\left(\frac{\overline\rho}{0.1\rho_{\rm drip}}\right)^{-1}\left(\frac{M_{\rm ej}}{10^{27}\,{\rm g}}\right)\,{\rm erg}.
\end{equation}
Here, $B$ is the local magnetic field near the star surface (which can greatly exceed the global dipole component that is probed by measurements of the star spindown rate). A large fraction of $E_{\rm flare}$ is carried by the ejecta Poynting flux. The asymptotic kinetic energy of the ejected baryonic blob $E_{\rm kin}\sim M_\text{ej}v_{\rm ej}^2/2$ should be a significant fraction of $E_{\rm flare}$. The radiated part (the energy dissipated by magnetic reconnection) is $E_\gamma\sim 0.1E_{\rm flare}$. These energetics are comparable to the parameters of the 2004 giant flare in SGR~1806-20, a significantly evolved magnetar with an estimated spindown age\footnote{\url{https://www.physics.mcgill.ca/~pulsar/magnetar/main.html}} $\sim 240$\,yr \citep{olausen_mcgill_2014}. Similar flares may occur more frequently in younger hyperactive magnetars. Eruptions with a broad range of $M_{\rm ej}$ are possible, depending on the magnetic configuration inside the neutron star. It can be quite different and more complex than assumed in our example simulation.

\subsection{Radioactive Transient}
\label{nuclear}
In addition to the prompt gamma-ray flare, the eruption will produce delayed gamma-rays, emitted by the ejected radioactive baryonic blob when it expands to gamma-ray transparency. Our simulation shows that most of the blob material was initially located rather deep in the crust, where hydrostatic pressure $g\Sigma$ is comparable to the magnetic pressure $B^2/8\pi$. Here, $g\sim 10^{14}$\,cm\,s$^{-2}$ is the gravitational acceleration, $\Sigma\sim \rho h$ is the column density, and $h\sim 10^4$\,cm is the hydrostatic scale-height of the crust. This gives a simple estimate for the initial density of the blob material,
\beq
  \rho\sim \rho_{\rm drip}B_{16}^2. 
\eeq
It is comparable to 
$10^{12}$\,g\,cm$^{-3}$ in our simulation. The initial nuclear composition of this material has the proton and neutron numbers $(Z,N)\sim (30,50)$. Before its advection to the surface by the diffusing magnetic loop, the material is squeezed by magnetic stresses. The blob then slowly emerges as an expanding mountain on the magnetar surface, with density decreasing to $\sim 0.1\rho_{\rm drip}$. Throughout this slow ambipolar drift phase, the blob remains thermally connected to the hot magnetar interior. As a result, at the onset of instability triggering the eruption, the elevated blob has temperature $T\sim 10^9$\,K, density $\rho\sim 0.1\rho_{\rm drip}$, and the electron fraction $Y_e\sim 0.4$.

The baryonic blob ejected during the eruption experiences fast expansion, on a timescale $r/v_{\rm ej}\sim 1\,r_7$\,ms. 
The sudden decompression lowers the electron Fermi energy and enables $\beta$-decays of the neutron-rich nuclei; the nuclei  also expel some neutrons. The heat produced by nuclear reactions temporarily offsets adiabatic cooling of the expanding blob (cf. the decompression $r$-process in initially cold ejecta described by \cite{Lattimer_1977} and \cite{Meyer_1989}). The ejected blob inevitably becomes radioactive, composed of nuclei with various lifetimes.

Strong radioactivity is a generic feature of massive crustal ejections, hot or cold. It produces observable nonthermal MeV photons when the blob expands for $\gtrsim 10^3$\,s and the ejecta becomes transparent to $\gamma$-rays \citep{Patel_2025}. By that time, the temperature of the  plasma and thermalized photons is strongly reduced by the adiabatic cooling. The drop of temperature below $10^5$\,K then leads to electron-ion recombination, and the baryonic blob releases its residual optical radiation from the radioactive energy thermalized during the opaque phase of expansion \citep{Patel_2025B}. Thus, our eruption model gives a robust prediction of delayed 
MeV and optical
flashes with a typical timescale $\gtrsim 10^3$\,s. Their 
light curves and spectra will require future calculations that 
employ nuclear reaction networks to trace the blob evolution from the onset of eruption to transparency.

\subsection{Fast Radio Burst}
\label{FRB}

The magnetosphere is significantly disturbed during the eruption, and this disturbance certainly involves a compressive component that launches an outgoing fast magnetosonic (FMS) wave. The wave has a characteristic frequency in the kHz band, which roughly corresponds to the eruption ms timescale. It expands through the magnetospheric plasma with the speed of light and forms a leading front of the eruption. Our simulation does not track such fast MHD dynamics, however we can make predictions based on recent studies of kHz waves launched in a neutron star  magnetosphere, including the prediction of a fast radio burst (FRB).

The recently obtained MHD solution for FMS waves expanding through the outer (dipole) magnetosphere shows the formation of monster shocks \citep{Beloborodov_2023}. On a microscopic level, shock formation is demonstrated by kinetic simulations that follow plasma particles \citep{Chen_2022,Vanthiegham_2025,Bernardi_2025}. The shock launched by the kHz FMS wave continues to expand beyond the magnetosphere into the wind zone and forms an ultra-relativistic blast wave of radius $r=ct$. 

Such a blast wave must emit a radio burst with a decreasing frequency, crossing the GHz band \citep{beloborodov_flaring_2017,Beloborodov_2020}; the emission frequency is set by the Larmor frequency of the plasma crossing the shock \citep{Hoshino_1992}.
The resulting FRB is relativistically beamed with a Lorentz factor $\Gamma>10^3$, which compresses its observed duration to $\lesssim 1$\,ms.
The mechanism of radio emission from shocks and its efficiency has been demonstrated in detail with kinetic plasma simulations \citep{Plotnikov_2019,Sironi_2021}. 

In addition, a radio burst may result from the interaction of the FMS pulse with the current sheet in the magnetar wind \citep{Lyubarsky_2020,Mahlmann_2022}. 
In this scenario,
the emission frequency 
depends on the power of the FMS pulse; it may or may not be in the GHz band.

Our expectation is that FRBs from shocks are a generic by-product of eruptions described in this Letter. The radio burst should arrive to the observer immediately preceding the gamma-ray flare. Its emission occurs within the solid angle occupied by the outgoing kHz FMS pulse generated by the eruption. We expect this angle to be large, with a significant chance to include the observer line of sight. However, no FRB was detected from the 2004 giant flare in SGR~1806-20 \citep{Tendulkar_2016}. For our model, this implies that the FRB was beamed away from the observer, and its 
solid angle was smaller than that of the gamma-ray emission.

\section{Discussion}

Analysis of
equilibrium 
magnetic configurations indicates
that neutron stars should contain mixed poloidal and toroidal magnetic fields \citep{flowers_evolution_1977,braithwaite_stable_2006}. In this Letter, we have 
shown
that the evolution of such configurations through ambipolar diffusion can lead to eruptions from the 
neutron star
surface, provided the internal field is very strong, $B> 10^{16}$~G. 
The toroidal field is essential for this mechanism because a purely poloidal field would contract under its own tension and dissipate inside the star, without emerging from the surface. 
The eruptions described in this work involve significant mass ejection, and 
are qualitatively different from other 
models 
in which the giant flares are triggered by the twisting of magnetospheric field lines 
\citep{thompson_giant_2001, Lyutikov_2006, Parfrey_2013}.

Young magnetars may contain substantially more complex magnetic fields than 
those considered in this work. If the internal field consists of many twisted flux ropes, they could rise to the surface through ambipolar diffusion and trigger a multitude of massive eruptions that form a nebula of decompressed crustal matter
around the magnetar. This process is efficient in the first $\sim 10^2$~yr when the high temperature $T\gtrsim 10^9$~K enables accelerated ambipolar diffusion.  
Such hyperactive magnetars are candidates for the engines of cosmological FRBs and their associated persistent radio sources (PRSs) \citep{beloborodov_flaring_2017,Rahaman_2025}. The single eruption in our simulation ejected enough matter to explain the number of particles ${\cal N}\sim 10^{52}$ needed for the PRS nebula around 
FRB 121102 \citep{Chatterjee_2017,beloborodov_flaring_2017}. 
The efficiency of mass ejection by more complex non-axisymmetric fields requires further investigation. 

Practically all known galactic magnetars are significantly evolved, with lower internal
temperatures $T\lesssim10^9$~K and slower ambipolar diffusion \citep{beloborodov_magnetar_2016}. Their 
current
activity may be largely driven by Hall drift, 
with rare eruptions due to strong remnant fields that were trapped in the crust at the end of the hyperactive phase. 
The galactic magnetars may have experienced
a period of hyperactivity and crustal 
eruptions
when they were young;
however, the nebulae 
created by earlier eruptions
would now be too faint to observe.
Only one galactic magnetar has an observed nebula, and its origin is consistent with an earlier period of intense flaring and outflows powered by 
ultra-strong internal fields \citep{younes_wind_2016,granot_learning_2016}. 

Our eruption mechanism may explain basic features of the 2004 giant flare and mass ejection from SGR 1806-20, possibly the youngest magnetar in our galaxy. It has an estimated spindown age $\sim 240$\,yr and the strongest magnetic dipole moment $\mu\approx 2\times 10^{33}$\,G\,cm$^3$ \citep{olausen_mcgill_2014}.
Although our simulation had 
$\mu\approx 2.8\times 10^{34}$\,G\,cm$^{3}$, 
eruptions will also occur when the external $\mu$ is weaker, provided the internal non-dipolar field is very strong, which may be the case for SGR 1806-20. As argued in this Letter, gamma-ray flares from such eruptions can be accompanied by a prompt FRB, delayed optical-UV and MeV flashes from the radioactive ejecta, and a radio afterglow. 

In this work, we demonstrated the eruption mechanism for a simple 
magnetic field configuration.
In the future,
it will be important to investigate the evolution of more complex and non-axisymmetric field topologies.
Diffusion
of ultra-strong 
internal
fields into the outer crust could trigger a range of buoyancy and interchange instabilities that produce smaller, more frequent eruptions. Modeling these instabilities will be essential to understand the eruption rates and energetics.

\section*{acknowledgments}
The authors thank Nicolas Chamel, Jonathan Granot, Dongzi Li, Anirudh Patel, and Chris Thompson for useful discussions. 
A.B. is supported by a PCTS fellowship and a Lyman Spitzer Jr. fellowship.
A.M.B. acknowledges support by 
NASA grants 21-ATP21-0056 and 80NSSC24K0282, NSF grant AST-2408199, and Simons Foundation grant No. 446228; this work was also facilitated by Multimessenger Plasma Physics Center (MPPC) grant PHY-2206609.
Y.L.'s work on this subject is supported Simons Collaboration on Extreme Electrodynamics of Compact Sources (SCEECS) and by Simons Investigator Grant 827103.

\appendix 

\section{Equations of Ambipolar Diffusion}
\label{setup}

\label{ambipolar_appendix}

We model the neutron star matter as $npe$ liquid with a baryon mass density $\rho$. It can also be viewed as a two-fluid system: the neutron fluid with number density $n_n$ and velocity $\vec{v}_n$, and the neutral plasma with density $n_p=n_e$ and velocity $\vec{v}_p=\vec{v}_e$. 
The  components $n$, $p$, and $e$ are each treated as an ideal degenerate Fermi gas, with chemical potentials $\mu_n$, $\mu_p$, and $\mu_e$, respectively.  This simplistic description also determines the pressure for each component and the hydrostatic structure of the star (which differs from the structure found with an accurate equation of state). It is sufficient for the purposes of this Letter, which focuses on basic features of ambipolar diffusion rather than the accurate hydrostatic structure.

Any deviation from chemical equilibrium $\Delta\mu = \mu_p + \mu_e - \mu_n\neq 0$ initiates $\beta$-reactions converting $p+e\rightarrow n$ with rate $\lambda\Delta\mu$ (negative $\Delta \mu$ drives the opposite conversion $n\rightarrow p+e$ with rate $-\lambda\Delta\mu$). Our calculation of $\Delta\mu$ is particularly crude in the lower crust, as we do not track its nuclear composition and effectively replace nuclei with protons. However, we find that the calculated evolution is insensitive to $\Delta\mu$, because it remains small ($\Delta\mu\ll kT$, see Appendix~\ref{beta_reactions}), and the contribution of charged baryons to pressure and total baryon density remains small.
In the outer crust, free neutrons no longer dominate, and all nucleons become locked into nuclei. Here, the nuclear composition remains unimportant because the system behaves according to single-fluid magnetohydrodynamics (MHD).

Equations expressing magnetic flux freezing in the plasma, continuity equations for the plasma and neutron components, and the thermal evolution of the neutron star can be written in form,
\begin{equation}
\frac{\partial \vec{B}}{\partial t} = \nabla\times(\vec{v}_p\times\vec{B}),
\label{Bfield}
\end{equation}
\begin{equation}
\frac{\partial n_p}{\partial t} + \nabla\cdot\left(n_p \vec{v}_p\right) =  -\lambda\Delta\mu,
\label{proton}
\end{equation}
\begin{equation}
\frac{\partial n_n}{\partial t} + \nabla\cdot(n_n\vec{v}_n) = \lambda \Delta\mu.
\label{neutron}
\end{equation}
\begin{equation}
    C_V\frac{d T}{dt } = - \dot{q}_\nu + \dot{q}_{amb},
\end{equation}
The thermal evolution equation tracks the internal temperature $T(t)$ of the neutron star, which controls the $p$-$n$ collision timescale $\tau_{pn}$, and the $p\leftrightarrow n$ conversion rate $\lambda$. The temperature profile is spatially uniform in our model because thermal conduction is fast compared to the timescales of interest. The heat capacity $C_V$ is that of degenerate neutrons, $\dot{q}_\nu$ is the volume averaged neutrino cooling rate due to murca reactions, and $\dot{q}_{amb}$ is the volume averaged ambipolar heating rate due to $p$-$n$ friction. Further details are described in \cite{beloborodov_magnetar_2016}.

In the core, $\lambda$ is set according to the modified urca reaction rates given by \cite{Sawyer_1989}. The value of $\lambda$ in the lower crust is given by the simplified model of $\beta$-reactions described in Appendix~\ref{beta_reactions}. In the upper crust and magnetosphere, $\lambda$ and the $n/p$ ratio are irrelevant, since the matter behaves as a single MHD fluid with a frozen $\vec{B}$.

Velocity fields $\vec{v}_p$ and $\vec{v}_n$ are described by equations stated below. The fluid motions are treated differently in the core, lower crust, upper crust, and magnetosphere. These four regions are defined in Section~\ref{numerical_setup}. We do not impose sharp boundaries between the different regions, but instead implement transition layers by smoothly interpolating the velocity fields from one region to the next as a function of local fluid variables. For example, the transition from core to crust is determined by the local baryon density, and the transition from outer crust to magnetosphere is determined by the local parameter $\beta=8\pi P/B^2$, where $P$ is the hydrostatic pressure in the star.

The equations for $\vec{v}_p$ and $\vec{v}_n$ involve gravitational potential $\Psi$.
It is prescribed by solving the Poisson equation in spherical symmetry, using the highly simplified equation of state (yielding a star of mass $M=1.11 ~M_\odot$). Although the outer layers of the star become strongly deformed from a spherical shape during the evolution, $\Psi$ is weakly affected because these layers contribute negligibly to the stars mass. Therefore, $\Psi$ is calculated once at the beginning of the simulation and used throughout.

In axisymmetry there are no pressure gradients to balance the azimuthal component of $\vec{j}\times\vec{B}/c$ (elastic forces are also negligible when the magnetic field is ultra-strong, Section~2.2). In order to satisfy the azimuthal force balance, we introduce an azimuthal velocity $v_\phi \propto (\vec{j}\times\vec{B})_\phi$ which ensures that $(\vec{j}\times\vec{B})_\phi\approx 0$ in the entire domain \citep{bransgrove_magnetic_2018}.

\subsection{Core}
In the core ($\rho>\rho_\text{core} \equiv 10^{14}$~g~cm$^{-3}$) we solve the multi-fluid ambipolar diffusion equations of \cite{Castillo_2020}. The proton velocity is 
\begin{equation}
    \vec{v}_p = \vec{v}_n + \vec{v}_\text{amb}
\end{equation}
where the ambipolar diffusion velocity is given by
\begin{equation}
\vec{v}_\text{amb} = \frac{\tau_{pn}}{m_p^\star n_p} \left[ \frac{\vec{j}\times\vec{B}}{c} -n_p\left( \nabla\mu_c  + \frac{\mu_c}{c^2}\nabla\Psi \right)  \right].
\end{equation}
Here $m_p^\star = \mu_p/c^2$ is the relativistic proton mass, $\mu_c = \mu_p + \mu_e$, $\Psi$ is the gravitational potential, and $\tau_{pn}$ is the proton-neutron collision timescale \citep{yakovlev_1990,Baiko_2001}. 

Following \cite{hoyos_asymptotic_2010}
and
\cite{castillo_magnetic_2017,Castillo_2020}, the neutron velocity is calculated from the condition of hydrostatic equilibrium of the total fluid
\begin{equation}
\vec{v}_n = \frac{1}{\rho\zeta}\left[ \frac{\vec{j}\times\vec{B}}{c} - n_p\left( \nabla\mu_c  + \frac{\mu_c}{c^2}\nabla\Psi\right) - n_n\left( \nabla\mu_n  + \frac{\mu_n}{c^2}\nabla\Psi \right) \right].
\label{force}
\end{equation} 
Here $\zeta$ is a relaxation parameter. It is introduced intentionally to create an artificial drag force $-\zeta\rho\vec{v}_n$, which is small compared to physical forces, so it weakly affects the physical force balance. This term replaces the inertial term $-\rho\partial_t\vec{v}$ (which would be present in the full MHD equations) and thus allows one to simulate the gradual evolution tracking the force balance with no need to resolve the short dynamical timescale. 
The choice of $\zeta$ weakly affects the results when the relaxation timescale $t_\zeta \sim R_\star/v_n$ satisfies $t_\zeta\ll t_{\rm amb}\sim 10^9$\,s. We use $\zeta^{-1}=10^{-16}$~s which implies $t_\zeta/ t_{\rm amb} \sim 10^{-2}$.  We have verified that increasing or decreasing $\zeta$ by a factor of 2 does not affect the evolution, as expected.

\subsection{Lower Crust}
Since neutrons are superfluid in the lower crust, they are not collisionally coupled to the crustal plasma, and each fluid must satisfy its own force balance. This gives:
\begin{equation}
 \vec{v}_p = \frac{1}{\rho_p \zeta_p}\left[ \frac{\vec{j}\times\vec{B}}{c} - n_p\left( \nabla\mu_c  + \frac{\mu_c}{c^2}\nabla\Psi \right)\right],
 \label{plasma_force}
\end{equation}
\begin{equation}
  \vec{v}_n  = - \frac{n_n}{\rho_n \zeta_n}\left( \nabla\mu_n  + \frac{\mu_n}{c^2}\nabla\Psi \right).
  \label{neutron_force}
\end{equation}
 We set the neutron relaxation parameter $\zeta_n=\zeta$ so that the neutron fluid gradually evolves through a slowly changing force balance and always remains close to hydrostatic equilibrium. The parameter $\zeta_p$ controls the timescale for the magnetized plasma in the crust to relax into a force equilibrium where $\vec{j}\times\vec{B}/c$ is balanced by pressure gradients and gravity (elastic forces are negligible when the magnetic field is ultra-strong, Section~2.2). We set $\zeta_p^{-1} = \tau_{pn}$, which is sufficiently small that the crust remains in equilibrium, and does not develop unbalanced forces. We set the rate of $\beta$-reactions $\lambda$ in the crust according to Eq.~(\ref{lambda_num}). Although we do not track the different nuclear species produced by $\beta$-reactions, we do capture the depletion of electrons and the associated reduction of pressure that enables slow compressions of the crustal matter. 

\subsection{Outer Crust and Magnetosphere}
Ambipolar diffusion cannot operate in the outer crust where neutrons are locked into nuclei. Therefore, the outer crust with ($\rho < \rho_\text{drip}$) is modeled as a single ideal MHD liquid. All particle species are advected together with the same velocity given by Eq.~(\ref{force}). We define the magnetosphere as the extreme low density regions where the plasma $\beta$ becomes smaller than a threshold value $\beta<\beta_0$. In this work we set $\beta_0 = 0.1$, however our results are not sensitive to the precise value as long as $\beta_0 \ll 1$. In the magnetosphere, we use magnetofrictional relaxation to evolve the magnetic field through a sequence of force free equilibria \citep{yang_force-free_1986}. In this method, field lines and fluid are advected together with velocity $\vec{v} \propto \vec{j}\times \vec{B}/|\vec{B}|^2$, so that $\vec{j}\times\vec{B}$ remains small at all times. The velocity of fluid parallel to field lines is given by the component of Eq.~(\ref{force}) that is parallel to $\vec{B}$.

\section{Numerical Method and Initial Conditions}
\label{appendix_numerical}

The equations in Appendix~\ref{ambipolar_appendix} are solved on a spherical $(r,\theta)$ grid using a specialized finite volume code. In order to preserve $\nabla\cdot\vec{B}=0$ the magnetic field is represented as $\vec{B} = \nabla\psi\times \nabla\phi + I\nabla\phi$, where $\psi$ is the flux function and $\nabla\phi = \hat{\phi}/ (r\sin\theta)$ [e.g. \cite{bransgrove_magnetic_2018}]. All variables are defined at the cell centers. The conserved variables $n_n$, $n_p$, and $I$ are solved with a conservative single-step finite volume method to achieve second-order accuracy in space and time \citep{Leveque_2001}.  The flux function $\psi$ satisfies an advection equation that is solved using an unstaggered single-step method that achieves second-order accuracy in space and time \citep{rossmanith_unstaggered_2006}. We use second-order Strang splitting to integrate the source terms in \Eqs~(\ref{proton}) and (\ref{neutron}). We use a specialized reconstruction technique to calculate the densities $n_p$ and $n_n$ at the cell edges and evaluate the gravitational source terms \citep{Krause_2019}. This ensures that hydrostatic equilibrium is not destroyed by spurious numerical errors. In the absence of magnetic fields our scheme preserves hydrostatic equilibrium to machine precision. 

The initial poloidal flux function is given by 
\begin{equation}
    \psi (r,\theta) =  
    \begin{cases}
    \displaystyle{
\psi_0\left[ \frac{35}{8}\left( \frac{r}{r_0}\right)^2 - \frac{21}{4}\left( \frac{r}{r_0}\right)^4 + \frac{15}{8}\left( \frac{r}{r_0}\right)^6\right]\sin^2\theta,
  }
  & (r<r_0)\\
    \displaystyle{
\psi_0 \left(\frac{r_0}{r}\right) \sin^2\theta,
  }
  & (r\geq r_0),
\end{cases}
\end{equation}
where we set the constants $r_0 = 1.05\times 10^6 $~cm and $\psi_0 = 2.7 \times 10^{28}$~G~cm$^2$. This implies the dipole moment $\mu = \psi_0 r_0\approx 2.8\times 10^{34}$~G~cm$^{3}$, and the dipole magnetic field strength at the pole $B_{\rm dip}=2\psi_0/r_0^2 \approx 4.8 \times 10^{16}$~G. The initial toroidal scalar function satisfies $I = I(\psi)$, so that the azimuthal component of the Ampere force vanishes. The precise form we use is
\begin{equation}
    I(r,\theta) = \begin{cases}
    \displaystyle{
I_0 \left(1 - \frac{\psi}{\psi_1}\right), }
& (\psi\geq\psi_1)\\
0, & (\psi< \psi_1),
\end{cases}
\end{equation}
where we set the constants $\psi_1 = 0.9 \psi_0$ and $I_0 = 1.5\times 10^{23}$~G~cm. This results in a toroidal field with maximum strength $B_{\phi,\rm max}\approx 5\times 10^{16}$~G.

\section{$\beta$-Reactions in the Crust}
\label{beta_reactions}
Let us consider a small volume of crustal matter that is initially in chemical $\beta$-equilibrium. Suppose the crustal plasma undergoes a small compression (or decompression) by the Ampere forces.
Note that the free neutron component is not compressed, as the Ampere force is applied to the plasma only, and the neutron superfluid flows easily through the lattice. Therefore, compression leads to an excess chemical potential of electrons that favors electron capture,
lowering the nuclei charge numbers $Z$. Note also that the inner crust has a high temperature $kT\approx 0.1$\,MeV, and a broad distribution of $Z$ is expected even for a fully relaxed composition \citep{Carreau_2020}; unlike cold catalyzed matter, it is weakly affected by the quantum shell effects. We leave calculations of the evolution of nuclear composition in the compressed hot crust to future work. Here, we emphasize the simple fact that compression dictates electron capture, as this allows the crustal plasma to keep near chemical equilibrium with free neutrons, whose Fermi level is unchanged by compression. As a result, the compressed region loses protons and electrons, allowing ambipolar drift.

This relaxation may be illustrated with a simplified model of a crust that contains only two nuclear species and a degenerate relativistic electron gas 
(we comment more on the effect of free neutrons and lattice corrections at the end of this section). In the simplified model, the allowed reactions are 
\begin{equation}
    (Z,A) + e^- \longrightarrow (Z-1,A) + \nu_e,
\end{equation}
\begin{equation}
    (Z-1,A) \longrightarrow (Z,A) + e^- + \bar{\nu}_e.
\end{equation}
We define the departure from local chemical $\beta$-equilibrium as
\begin{equation}
    \Delta \mu \equiv  E_F - Q,
\end{equation}
where $E_F$ is the electron Fermi energy and $Q = E\{ Z-1,A\} - E\{ Z,A\} + m_e c^2$ is the threshold for the electron capture reaction.
We denote the number density of species $(Z,A)$ and $(Z-1,A)$ by $n_+$ and $n_-$, respectively. The electron density is $n_e = a_\mu ([E_F/m_e c^2]^2 - 1)^{3/2}$, where 
$a_\mu = (3\pi^2 \lambdabar _e^3)^{-1}$
and $\lambdabar _e = \hbar/(m_e c)$. The net volumetric rate of $(Z,A)$ production due to electron captures and $\beta$-decays is given by 
\begin{equation}
    \dot{n}_+ = \Gamma_+ = n_-\lambda_- - n_+\lambda_+,
\end{equation}
where $\lambda_+$ is the electron capture rate per $(Z,A)$ nucleus and $\lambda_-$ is the $\beta$-decay rate per $(Z-1,A)$ nucleus (\cite{Tsuruta_1970}, see their \Eq~(14)
and Appendix~1). The net rate of $(Z-1,A)$ production is $\dot{n}_- = \Gamma_- = -\Gamma_+$. 

Chemical $\beta$-equilibrium is defined by $\Gamma_+ =\Gamma_- = 0$ and $\Delta\mu=0$. 
Together with the condition for charge neutrality $n_e = Z n_+ + (Z-1)n_-$, this gives the equilibrium density of each species 
\begin{equation}
    n_\pm = n_e \frac{\lambda_\mp}{Z\lambda_- + (Z-1)\lambda_+},
\end{equation}
where the right hand side is evaluated at $\Delta\mu=0$. 

A small compression of the equilibrium matter raises the electron Fermi energy $E_F\longrightarrow Q + \Delta\mu$ and pushes the crust out of chemical equilibrium ($\Delta\mu >0$). This modifies the reaction rates $\lambda_\pm \longrightarrow \lambda_\pm + \delta \lambda_\pm$, which tend to drive the system back toward equilibrium. Retaining terms of linear order in $\Delta\mu$ and assuming $\Delta\mu \lesssim kT \ll Q$, we find the net reaction rate induced by the compression
\begin{equation}
    \Gamma_+ \approx n_- \delta\lambda_- - n_+\delta\lambda_+ \approx - \lambda \Delta\mu.
    \label{weak_rate}
\end{equation}
The coefficient $\lambda$ is related to the compressibility of the crustal plasma and is given by 
\begin{equation}
    \lambda = n_+ \langle F^+\rangle\frac{4\ln{(2)}}{ft} \frac{Q^2(kT)^2}{(m_e c^2)^5},
    \label{lambda}
\end{equation}
where $\langle F^+\rangle$ is a dimensionless factor that accounts for the Coulomb interaction of the electron and the nucleus \citep{Tsuruta_1970}, and $ft$ is the effective half-life of the nucleus. If we instead consider a small decompression of the crust ($\Delta\mu <0$), we find the same reaction rate, but in the opposite direction. Therefore, \Eqs~(\ref{weak_rate}) and (\ref{lambda}) apply for both signs of $\Delta\mu$. The $ft$ value of known nuclei spans $\log ft\sim 3-8$. Its value for nuclei in the deep crust is uncertain, but $ft \sim 10^6$~s is generally expected \citep{Lau_2012}. We estimate the reaction rate as
\begin{equation}
    \lambda \approx 4\times 10^{36}\,\rho_{14}~T_9^2\left( \frac{Y_e}{0.04}\right)\left( \frac{Q}{20~\text{MeV}}\right)^2 \left( \frac{ft}{10^6~\text{s}}\right)^{-1}~\text{~s}^{-1}\text{cm}^{-3}~\text{erg}^{-1},
    \label{lambda_num}
\end{equation}
with $Y_e$ is the electron fraction. We conservatively set $ft=10^8$~s in the simulation, and find that $\beta$-reactions are still fast enough for the crustal magnetic field to evolve on a timescale $t_{\rm amb}\sim 10^9$~s.  Lattice corrections and the presence of free neutrons at $\rho>\rho_\text{drip}$ modify the effective $Q$-value of the reaction \citep{Chamel_2015}. Studies with realistic nuclear reaction networks find that compressions of the crust result in a distribution of nuclear species \citep{Lau_2018}. A more sophisticated calculation is needed to investigate the changing distribution of nuclei produced by magnetic compressions at high temperatures, and this is left for a future dedicated work. In this work we focus on the magnetic field evolution, which only requires the depletion of electrons to enable compressions of the crustal plasma. The pressure of the ion lattice is negligible, regardless of which ion species are present.

\bibliographystyle{apj}

\end{document}